\documentclass[draft]{agujournal2019}
\usepackage{url} 
\usepackage{lineno}
\usepackage{soul}
\usepackage{xspace}

\newcommand{\cassini}{\emph{Cassini}\xspace}
\newcommand{\voyager}{\emph{Voyager}\xspace}
\newcommand{\corot}{\emph{CoRoT}\xspace}
\newcommand{\kepler}{\emph{Kepler}\xspace}
\newcommand{\tess}{\emph{TESS}\xspace}
\newcommand{\juno}{\emph{Juno}\xspace}

\journalname{AGU Advances}

\begin{document}

\title{Saturn's rings as a seismograph to probe Saturn's internal structure}

\authors{Christopher R. Mankovich}
\affiliation{1}{Division of Geological and Planetary Sciences, California Institute of Technology, Pasadena, CA 91125, USA}
\correspondingauthor{Chris Mankovich}{chkvch@caltech.edu}

\begin{keypoints}
\item \cassini characterized more than 20 waves in Saturn's rings caused by Saturn's oscillations, opening the door to giant planet seismology.
\item The frequency spectrum has revealed that Saturn's deep interior is stably stratified, and yields a seismological rotation rate for Saturn.
\item The existing data can quantify the location and strength of Saturn's deep stable stratification, as well as Saturn's differential rotation.
\end{keypoints}

\begin{abstract}
 As it has already done for Earth, the sun, and the stars, seismology has the potential to radically change the way the interiors of giant planets are studied. In a sequence of events foreseen by only a few, observations of Saturn's rings by the \cassini spacecraft have rapidly broken ground on giant planet seismology. Gravity directly couples the planet's normal mode oscillations to the orbits of ring particles, generating spiral waves whose frequencies encode Saturn's internal structure and rotation. These modes have revealed a stably stratified region near Saturn's center, and provided a new constraint on Saturn's rotation. 
\end{abstract}

\section*{Plain Language Summary}
Just like measuring earthquakes around the world can tell scientists about Earth's deep structure, vibrations of gas giant planets can tell us about their deep structure. But these vibrations are very hard to detect. At Saturn, help has come in the form of Saturn's icy rings, where gravity causes the orbits of ring material to pick up the planet's steady vibrations. This makes waves in the rings that are now being used as a powerful tool to study the inner workings of Saturn itself. Surprisingly, these waves have shown that the fluid motions in the deepest parts of the planet are relatively tame, compared to the forceful churning motions that were generally expected. They have also provided {a measurement} of the length of a Saturn day, a tough quantity to {determine}.

\section{Introduction}
The structure and makeup of the gas giants are key tracers of the planet formation process. Piecing together this ancient history demands answers to several entangled questions: {D}id the gas giants form around solid planetesimal cores? If so, to what extent do these cores survive the process that then delivers hydrogen and helium, the bulk of these planets’ mass? How are the heavier constituents like ice and rock distributed after formation, and redistributed during the subsequent billions of years of evolution? Are the gas giants convective throughout their interiors, as has usually been assumed?

Within just the past five years or so, the interior mass distributions and rotation profiles of Jupiter and Saturn have been better constrained than ever owing to up-close observations of their gravity fields by spacecraft like \juno \cite{2018Natur.555..220I,2018Natur.555..223K,2018Natur.555..227G} and \cassini \cite{2019Sci...364.2965I,2019ApJ...879...78M}. However, the gravity fields alone are largely insensitive to the greatest depths in these planets, where precious clues about the planet formation process lie hidden. In the case of Saturn, a totally independent means of peering into the planet’s interior is emerging thanks to information encoded in---of all places---Saturn's rings. 

Like any system in a stable equilibrium, planets respond to small perturbations by oscillating about that equilibrium state. Earth, for example, rings like a bell for days following a major earthquake. Global seismology deciphers the frequencies of these large-scale trapped waves---the normal modes of oscillation---to understand our planet's internal structure \cite{Dahlen1998}. 

The stars, too, vibrate. Helioseismology, the study of our sun’s trapped acoustic wave oscillations, has revealed most of the sun's internal rotation profile in detail as well as the depth of the solar convection zone \cite{1991ApJ...378..413C,1996Sci...272.1286C}. These oscillations are excited not by tectonics as on Earth, but by turbulent convection in the sun's outer layers, just one of several processes that causes stars to vibrate quite generally. Beyond the solar system, tens of thousands of stars from the main sequence through the red giant branch have had their interior oscillation frequencies measured from their rapid brightness variations through time. These data have provided entirely new information about the physics of stellar evolution, rotation, and internal heat transport, and yielded powerful handles on stellar parameters like density, surface gravity, age, and inclination that are vital to studies of exoplanet systems \cite{2013ARA&A..51..353C}. This field of asteroseismology---the study of stellar interiors using normal mode oscillations---has led to something of a renaissance in stellar astrophysics over the last 15 years as a result of space missions like \corot, \kepler, and now, \tess. 

In light of the major advances that normal mode seismology {has }brought to terrestrial, solar, and stellar physics over the last several decades, similar methods hold immense promise for revealing the unseen inner workings of giant planets. Efforts to detect trapped oscillations in the gas giants from ground-based telescopes have been underway for more than 30 years, focusing for the most part on Jupiter \cite{1989ApJ...343..456D,1991A&A...248..281S}. This is because Jupiter’s large angular size and lack of a prominent ring system obscuring its surface make it amenable to seismological study by Doppler imaging, wherein a time series of line-of-sight velocity maps of the planet’s rumbling surface reveal the trapped oscillations that are in turn examined in the frequency domain. These studies have so far culminated in an encouraging detection of excess power at mHz frequencies consistent with Jupiter’s trapped acoustic waves \cite{2011A&A...531A.104G}. However, the isolation of individual normal mode frequencies---a necessary step to connect measured frequencies with knowledge of the planet’s interior---is stymied by the level of noise in the data gathered so far. Longer continuous coverage provided by observations from several longitudes on Earth may bring ground-based acoustic mode seismology of Jupiter within reach in the coming years \cite{2013ASPC..478..119S}. In the meantime a very different, and ultimately complementary, method for studying giant planet oscillations has come to light thanks to \cassini's campaign at Saturn.

\section{Kronoseismology}
The very rings that so inconveniently obscure part of Saturn’s disk on the sky turn out to offer the so far singular window into the individual normal-mode oscillations of a giant planet. Confirming a decades-old hypothesis \cite{stevenson1982} and a pioneering body of theoretical work that followed \cite{1990PhDT.........3M,1991Icar...94..420M,1993Icar..106..508M}, NASA’s \cassini mission to Saturn has decisively shown that the periodic variations in Saturn’s gravity field caused by the planet’s internal oscillations in turn disturb the typically well-ordered orbits of particles in Saturn’s icy rings \cite{2013AJ....146...12H,2014MNRAS.444.1369H,2016Icar..279...62F,2019Icar..319..599F,2019AJ....157...18H}. This regular forcing stirs up waves that are wound into spiral patterns by the rings’ differential rotation---the same process by which a rotating bar structure in the center of a galaxy can organize the stellar, gas, and dust mass into spiral arms. A key difference in Saturn’s rings is that there the waves are very tightly wound around the planet, a result of Saturn’s immense mass compared to the mass in the rings themselves. As a result, the radial wavelength of these waves is of order a mere kilometer, versus the whopping 70,000 km scale of the main rings overall. The effect of the waves is therefore invisible from afar; their detection requires an up-close view the likes of which only a spacecraft mission can provide. 

\begin{figure}
  \centering\includegraphics[width=1.0\textwidth]{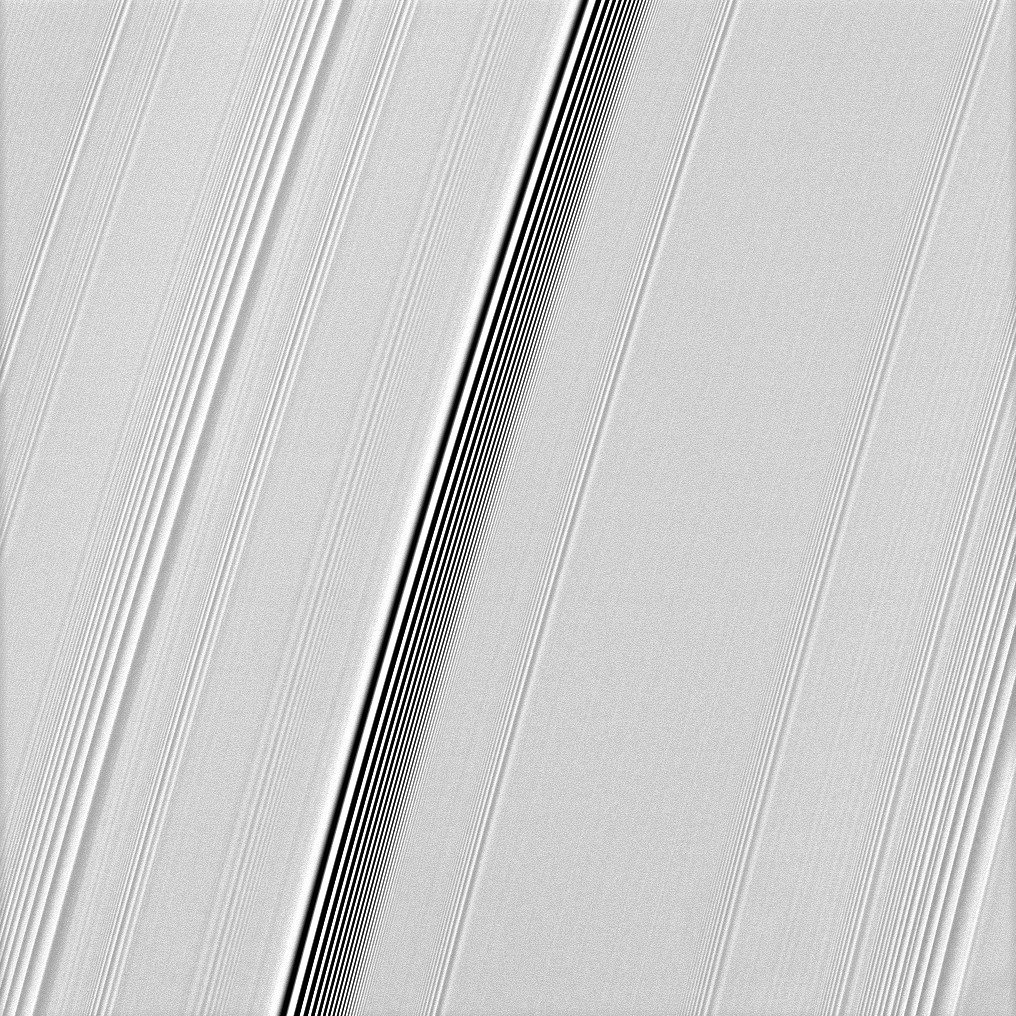}
  \caption{A strong spiral bending wave (near center, with wavelength decreasing toward Saturn, to the right in this image) and several weaker spiral density waves (wavelength decreasing away from Saturn) as observed in Saturn's A ring by \cassini's Imaging Science Subsystem narrow angle camera. Credit: NASA/JPL/Space Science Institute; retrieved from \url{https://photojournal.jpl.nasa.gov/catalog/PIA12545}.
  }
\label{fig.waves}
\end{figure}

\begin{figure}
\centering\includegraphics[width=0.5\textwidth]{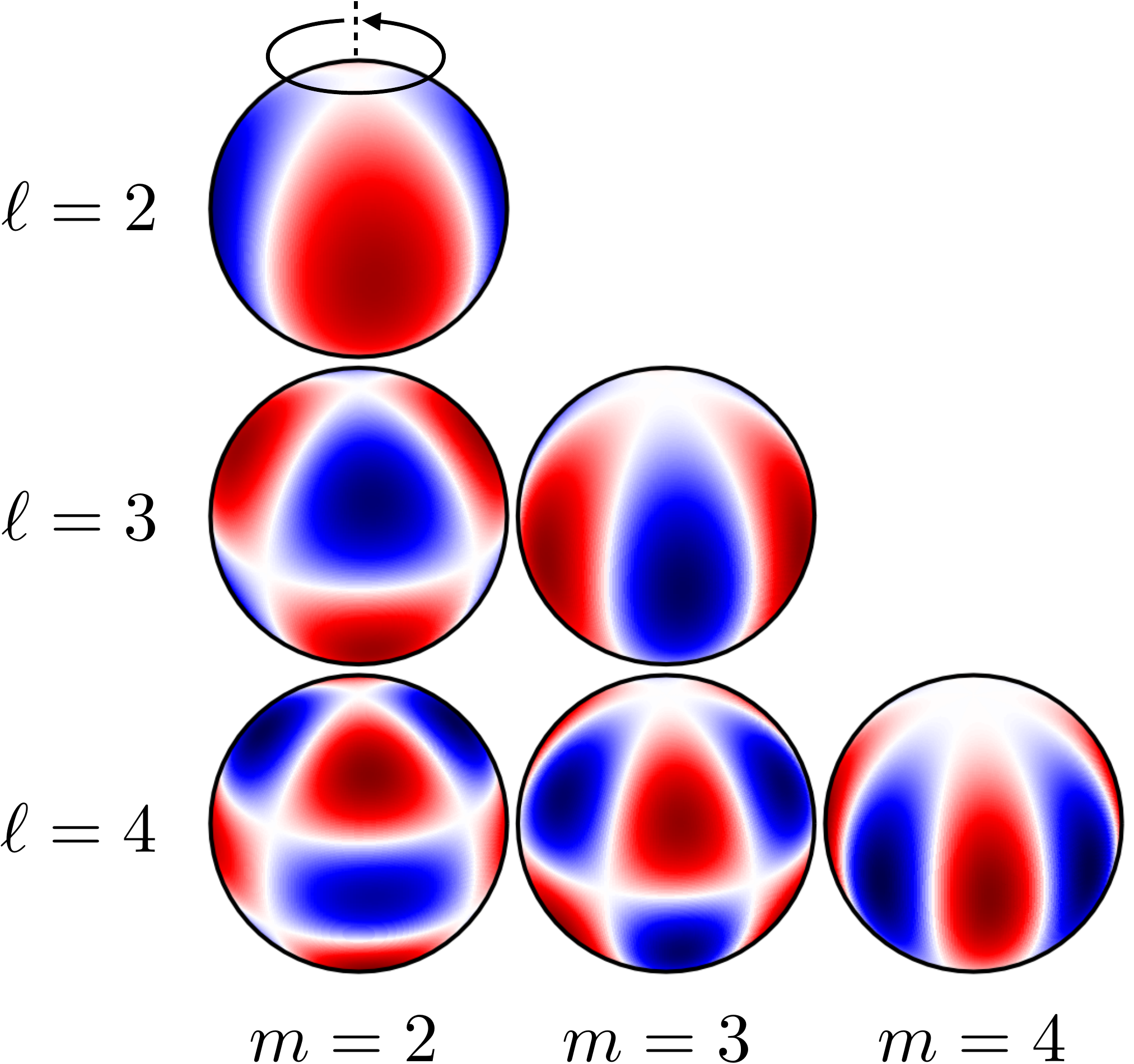}
\caption{A visualization of some of the spherical harmonics relevant for Saturn ring seismology, labeled by their angular degree ($\ell$) and azimuthal order ($m$). The color map corresponds to the magnitude of the perturbations---e.g., to the density and gravity field---as a result of the oscillation. An inertial observer sees each $m\neq0$ pattern rotating as a function time, the combined effect of the planet's rotation and the steady propagation of the wave pattern around the planet.}
\label{fig.harmonics}
\end{figure}

Spiral waves in Saturn's rings were first studied intensely during the \voyager era, when it became clear that periodic gravitational perturbations from Saturn's satellites launch an abundance of spiral waves throughout the rings \cite{1981Natur.292..703C,1983Icar...53..185S}. 
Each wave falls into one of two classes: density waves are alternating compressions and rarefactions of orbits confined to the ring plane, whereas bending waves are alternating vertical departures above and below the ring plane.
{A given satellite orbit generally gives rise to both types of wave, although only inclined satellites can drive bending waves.}
{Figure~}\ref{fig.waves}{ displays some examples of waves excited by gravitational forcing by satellites.}
The physical description of the ring response to this slow periodic forcing by satellites applies equally well to the faster forcing by normal mode oscillations inside Saturn, which indeed create their own density and bending waves in the rings.
{These global planetary oscillations take place at countless individual frequencies, their overall spectrum dictated principally by the planet's mean density, its compressibility as a function of depth, its rotation, and interfaces or gradients in its chemical composition} \cite{1989nos..book.....U}{.}
In the language of the spherical harmonics---a convenient language for separating the {frequency} components of the complicated overall planet oscillation (Figure~\ref{fig.harmonics})---modes with even $\ell-m$ {induce radial oscillations in the orbits of ring particles, driving density waves. Modes with odd $\ell-m$ on the other hand induce vertical oscillations in ring orbits, driving bending waves} \cite{1993Icar..106..508M}.

However, while Saturn vibrates, most parts of the rings experience a negligible response. For a given oscillation {mode} in the planet, the {frequency of the gravitational forcing experienced by an orbiting ring particle depends on the oscillation frequency of the mode, its azimuthal order $m$, and the orbital frequency of the particle.}
{The} vast majority of ring orbits couple to {the mode} quite poorly because ring particles will experience extrema of the planet oscillation at random orbital phases; the forcing tends to cancel, and no coherent response can develop. 
But {at the special radial locations in the ring at which the forcing frequency coincides with the orbital frequency,} each extremum in the planet oscillation forcing takes place at about the same orbital phase, and a coherent response will develop. This is the condition of resonance: a commensurability of the forcing planet frequency and the natural frequency of a ring orbit. Ring seismology is thus sensitive only to the range {of }frequencies that are occupied by ring orbits, setting intrinsic limits on the type of oscillation within Saturn that this method can probe.  

Because the frequencies of ring orbits decrease steeply with distance from Saturn, distinct planet oscillation modes excite waves at distinct locations in the rings. This means that when these waves can be detected, they are spatially separated according to the frequency and geometry ($m$ value) of the corresponding normal mode in Saturn. Saturn’s rings thus, incredibly, form a natural frequency-domain seismograph for the planet’s normal mode oscillations. 

\begin{figure}
\centering\includegraphics[width=\textwidth]{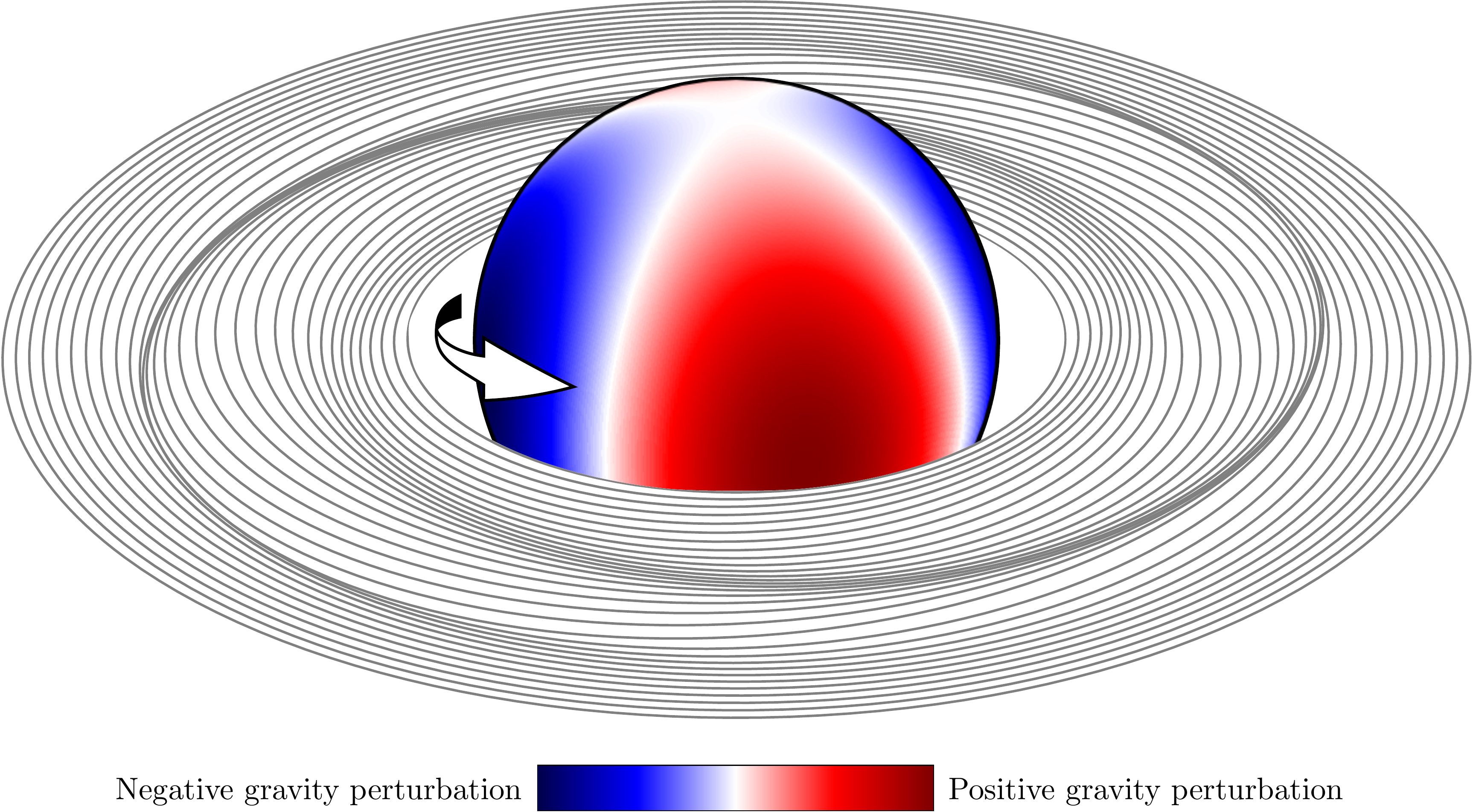}
\caption{A schematic of an $\ell=2$, $m=2$ normal mode of oscillation inside Saturn generating a two-armed spiral density wave in the rings. In reality spiral patterns in the rings are much more tightly wound, and are only evident near a resonance.}
\label{fig.mode_wave_schematic}
\end{figure}

\cassini was able to realize these ideas by peering through Saturn’s rings toward bright stars and recording the variation in transmitted light as the spacecraft moved in its orbit. As the line of sight passes through a wave in a translucent part of the rings, the transmitted starlight varies sinusoidally, and the wave pattern can be reconstructed to obtain the precise location of the resonance and thus the frequency of the perturbing planet mode. Furthermore, by making repeated passes as \cassini orbited Saturn for longer than a decade, scientists have been able to observe each wave from multiple perspectives. This broader view allowed them to count the number of spiral arms in each spiral wave pattern, a crucial piece of information for discriminating which mode of the planet’s oscillation is responsible. (An $m=2$ mode in Saturn creates a two-armed spiral, an $m=3$ mode a three-armed spiral, and so on; see Figure \ref{fig.mode_wave_schematic}.) A spate of recent \cassini results \cite{2013AJ....146...12H,2014MNRAS.444.1369H,2016Icar..279...62F,2019Icar..319..599F,2019AJ....157...18H} has characterized about two dozen spiral waves associated with normal mode oscillations inside Saturn, providing for the first time a power spectrum suitable for normal mode seismology of a giant planet. Hedman, Nicholson and their collaborators termed this field Kronoseismology, after the Greek name for Saturn. As it turns out, even as the waves that emerged from these data validated the hypothesis of the rings as a natural seismograph, they also revealed surprises of profound consequence for studying Saturn’s interior. 

\subsection{Deep interior structure}
The expectation from Marley and Porco’s theory was that ring waves would be seen at resonances with Saturn’s fundamental mode oscillations, i.e., {standing} surface gravity waves. {These modes are fundamental modes in the sense that they have no nodes as a function of radius inside the planet; in terrestrial seismology they correspond to the fundamental spheroidal modes.}
{Marley and Porco} showed that these resonances would lie almost entirely in an inner region of the rings known as the C ring, a fortuitous alignment because the translucent C ring transmits enough starlight to make these experiments possible. (The heftier A and B rings that dominate the rings’ visual appearance are generally opaque to starlight.) They predicted an ordered pattern of resonances at distinct locations, and that the normal mode of Saturn responsible for each observed wave feature would be readily apparent based on the observed number of spiral arms. Instead, what Hedman and Nicholson discovered were \emph{clusters} of waves (a pair of $m=2$ waves; a triplet of $m=3$ waves) in the proximity of the strongest fundamental mode resonances, an impossibility if the detailed model that Marley and Porco had proposed 20 years earlier represented the whole truth. What the data showed was unambiguous; what they demanded was a reexamining of the assumptions that had been made so far about the physics at work in Saturn's interior.

The origin of these unexpected waves did not stay mysterious for long: it was soon demonstrated that they could be naturally produced if Saturn’s interior hosts not only the expected fundamental modes, but also gravity modes---trapped internal gravity waves \cite{2014Icar..242..283F}. 

The implication that Saturn supports internal gravity waves is profound because their presence requires part of Saturn’s fluid interior to be stably stratified, a stark departure from the common assumption that Saturn’s interior is fully convective. A stable stratification means that a vertically displaced fluid parcel will tend to return to its starting position, enabling oscillations{ at a characteristic (Brunt-V\"ais\"al\"a) frequency determined by the gravity, density gradient, and compressibility}. By contrast, in a convective environment, a similarly displaced fluid parcel would simply continue to accelerate away from its starting position, so that no periodic fluid motion could be sustained. 

This stable stratification suggests that Saturn’s deep interior has a significant composition gradient wherein molecular weight increases toward the planet’s center, mitigating the unstable temperature gradient that if left to its own devices would trigger convection and large-scale mixing of material. Instead, the gravity modes suggest a relatively quiet, extended, smooth transition between a dense rock- and ice-dominated core and the less dense hydrogen-dominated envelope. 

While Fuller presented strong evidence that the mixture of the fundamental modes and gravity modes was responsible for the complicated spectrum of waves observed in the rings up to that point, the model was effectively a proof of concept: the ideas have yet to be {turned} into quantitative knowledge of Saturn’s deep interior. Updated analyses that address the Saturn-associated ring waves discovered in more recent years---and that apply more detailed and realistic models for Saturn’s interior structure---will offer meaningful constraints on the location and extent of Saturn’s deep stable stratification. Because of the sensitivity of these waves to the deepest regions inside Saturn, these new constraints will serve as an invaluable complement to the gravity science \cite{2019ApJ...879...78M,2019GeoRL..46..616G} that has come out of the end of the \cassini mission. 

\subsection{Rotation}
The second major advance to come {from} ring seismology is the window it offers into Saturn’s interior rotation. One of the major historical unknowns about the Saturnian system is just how quickly Saturn rotates, a quantity of fundamental importance but one that is difficult to measure. Meteorological features can be tracked as Saturn rotates, but as on Jupiter or Earth, flows associated with the weather do not track the rotation of the bulk of the planet’s mass. Even among planets with no solid surface, Saturn’s rotation is exceptionally difficult to pin down. 
The virtually perfect alignment of its magnetic dipole axis with its rotation axis \cite{CAO2019113541} means that no obvious trace of the planet’s rotation is visible from afar; this stands in contrast to Jupiter, where the rotating magnetic field produces a strong periodic radio emission that is ideal for tracking the planet’s spin. As a result, Jupiter’s spin period has long been known to the level of milliseconds \cite{1960AJ.....65S.487D}, while estimates for Saturn’s spin period historically vary {between roughly} 10 hours 30 minutes {and} 10 hours 50 minutes \cite{1981GeoRL...8..253D,2006Natur.441...62G,2007Sci...317.1384A}. While this spread is only a few percent of a Saturn day, it significantly muddies the waters when it comes to understanding Saturn's atmospheric and interior flows, its overall interior structure, and consequently the formation and evolution pathway that Saturn has undergone. With this historical challenge in mind, \cassini was tasked with finding new means of constraining Saturn’s interior rotation. Indeed, unexpectedly, Saturn ring seismology has proven to be one such path. 

Putting aside the subset of C ring waves complicated by the mixture of fundamental and gravity mode oscillations, the remainder of Saturn-associated waves detected to date---14 out of a total of 21---are well understood as resonances with simple fundamental mode oscillations of Saturn, the likes of which Marley and Porco had anticipated. These planet modes have higher frequencies and angular degrees $\ell$, and are consequently confined somewhat closer to the surface, diminishing their value for constraining the structure of Saturn’s deep interior. However, there is a tradeoff at play: these shallower, higher-$\ell$ modes also intrinsically possess more angular structure, and as a result are dramatically more sensitive to Saturn's rotation. Detailed calculations show that even accounting for the significant uncertainties in modeling Saturn's interior structure, the quality of fit to the ensemble of ring wave frequencies is dominated by the rotation rate assumed for Saturn's interior. To leading order this sensitivity is due to the Doppler shift relating a frequency in the planet's rotating reference frame to the inertial reference frame appropriate for studying the ring response. Rotation also subtly modifies mode frequencies by inducing Coriolis forces and rendering the planet oblate, adding significant complexity to the frequency calculation.
This sensitivity forms a basis for a recent seismological measurement of Saturn's rotation rate \cite{2019ApJ...871....1M}. The resulting period{ of $10{\rm h}\, 33{\rm m}\, 38{\rm s}^{+1{\rm m}\, 52{\rm s}}_{-1{\rm m}\, 19{\rm s}}$} is fast compared to radiometric and magnetic periods observed by spacecraft and long used as a proxy for the planet's interior rotation \cite{1981GeoRL...8..253D}; the faster seismological estimate is instead consistent with recent estimates based on Saturn's shape and gravity field \cite{2015Natur.520..202H,2019ApJ...879...78M} and the stability of its jet streams \cite{2009Natur.460..608R}, strengthening the growing consensus that periodic modulations associated with Saturn's magnetosphere are not well coupled to the rotation of Saturn itself \cite{2007Sci...316..442G}.

The full power of the seismological probe of Saturn's rotation has yet to be {realized}, however. In contrast to the extraordinarily precise frequencies provided by ring seismology, the theoretical methods employed so far to predict mode frequencies from an interior structure model are significantly imprecise as a result of their approximate treatment of rotation effects. Even with perfect knowledge of Saturn's interior structure, these methods can only predict fundamental mode frequencies with a relative precision of order $10^{-3}$ at best; by comparison the observations by \cassini yield wave frequencies with a typical relative precision of $10^{-5}$. In particular, the seismology delivers Saturn's rotation period to a precision of about 1.5 minutes, an uncertainty comparable with the more model-dependent constraints based on Saturn's shape and gravity field, but significantly larger than that derived from the stability of atmospheric flows. Whether the seismology, gravity-shape, and atmospheric dynamics constraints will converge on a consistent rate for Saturn's bulk rotation thus awaits improved theoretical methods for the seismological forwarding modeling; these will take the form of {either} higher-order asymptotic treatments of rotation \cite{1998A&A...334..911S,2008ChJAA...8..285K}, or {non-perturbative methods that can treat rotation free of approximations} \cite{2006A&A...455..621R,2012A&A...547A..75O,2017PhRvD..96h3005X}. Because the {rotation contributions to the} fundamental mode frequencies scale linearly with Saturn's rotation rate{ to leading order}, if the theory can match the data at a relative precision of $10^{-5}$, the existing seismology data could in principle yield Saturn's rotation period to within a second. In reality, at this level, matters are complicated by \emph{differential} rotation: rather than measuring any single rotation rate, it is more appropriate to speak of quantifying Saturn's rotation \emph{profile}. {Generally speaking the rotation rate in fluid planets may with both depth and latitude, as is known to be the case in the Sun on the basis of helioseismology }\cite{1989ApJ...343..526B,1991ApJ...367..649G}.

The discovery of deep differential rotation in Saturn was a major advance to come out of \cassini gravity science \cite{2019Sci...364.2965I,2019GeoRL..46..616G}, echoing a similar discovery at Jupiter by the \juno spacecraft reported only months earlier \cite{2018Natur.555..223K,2018Natur.555..227G}. It had been understood for some time that the electrically conductive deep interiors of both planets---for Saturn, roughly the inner half by radius---should be kept rigidly rotating by electromagnetic forces. What has not become clear until recently is how the interior flows are organized between that rigid fluid metallic interior and the east-west zonal flows apparent on Saturn's surface: are the surface flows a shallow atmospheric phenomenon, or are they deep-seated? Structure in Saturn's gravity field as observed at the end of the \cassini mission has rapidly shed light on this question, showing that the east-west zonal winds evident on Saturn's surface indeed penetrate to significant depth in the interior, to approximately 9,000 km---15\% of the planet's radius---below the surface \cite{2019Sci...364.2965I,2019GeoRL..46..616G}. Such a deep flow pattern must indelibly alter the frequencies of the fundamental mode oscillations, an effect studied by \citeA{1993Icar..106..508M} but one that has yet to be considered in the detailed numerical calculations used to interpret the glut of mode frequencies now available. Notably, the fundamental modes present in the data have angular degrees covering almost all values from $\ell=2$ to $\ell=14$, meaning that they probe a wide range of depths in Saturn and thus, when taken together, offer a sensitive handle on the differential rotation. Realizing this potential will require the kind of theoretical improvements described above to accurately account for Saturn's rapid rotation, an endeavor that will enable an independent confirmation of the rotation profiles derived from gravity science. Of course, in pursuing this brand-new line of observational evidence, there is also the potential to uncover surprises.

\section{Conclusions \& Outlook}
The frequencies of 21 normal modes of oscillation in Saturn have been measured from waves in high-resolution profiles of ring-occulted starlight. 

Seven of these modes (those with $m=2$ and $m=3$) appear to be rooted in mixed gravity-fundamental modes. Their gravity mode character requires that a significant fraction of Saturn's deep interior---potentially most of the inner half by radius---is stabilized against convection by composition gradients. {This echoes the evidence for a dilute core structure in Jupiter from \emph{Juno} gravity science} \cite{2017GeoRL..44.4649W}{, although on the basis of gravity data alone it's unclear whether that signal comes from a continuous composition gradient (stable stratification) or from a uniformly enriched region (still fully convective).}
The detection of {mixed} modes{ in Saturn} is the strongest evidence to date that the {Saturn's }fluid envelope is not fully convective, {a general conclusion supported by} independent indications from {Saturn's magnetic moments measured by \emph{Cassini}} \cite{CAO2019113541}. {However, the deep, relatively thick stable stratification suggested by the mixed mode seismology poses something of a challenge for models of Saturn's magnetic field generation, which to date have appealed to the fundamentally different picture of a deep fully convective dynamo region surrounded by only a thin (5-15\% of Saturn's radius) stably stratified shell} \cite{1982GApFD..21..113S,doi:10.1029/2009GL041752,2016PEPI..250...31S,CAO2019113541}. {Any stable stratification inside Saturn is in fact likely to undergo double-diffusive convection }\cite{2013NatGe...6..347L}{, and it remains possible that the associated weakly turbulent motions could play a role in Saturn's dynamo.}

This {seismological evidence for a stable stratification in Saturn} fundamentally alters the picture of the planet's deep interior. {The mixed modes resonating with the rings} are the most direct probes of the deepest inner workings of Saturn yet available, and a quantitative understanding of the deep distributions of hydrogen, helium, rocks and ices---of central importance to formation models---awaits the systematic application of more realistic interior models to the seismology data.

The remaining modes (those with $m\geq4$) correspond to pure fundamental modes. They carry less information about Saturn's interior structure and more about its rotation profile, allowing the first seismological measurement of Saturn's bulk rotation rate. The current data will provide stringent constraints on differential rotation within Saturn, but only after the theory is extended to more accurately treat Saturn's rapid rotation, including its dependence on depth and latitude within the planet.

{Saturn's rings are an unparalleled tool for sounding the inside of a giant planet, but can ring seismology be applied elsewhere? The tenuous, dusty ring systems around Jupiter and Neptune, for example, seem inauspicious for ring seismology of the kind described here. The best candidate for ring seismology beyond Saturn is likely Uranus, whose richer ring system includes the $\gamma$ ring, a feature apparently undergoing forcing of unknown origin} \cite{1986Sci...231..480F}{.}

Summarizing, the current moment leaves a few important gaps to be bridged{ at Saturn}: 
\begin{enumerate}
  \item Theoretical Saturn mode frequencies computed so far are imprecise, while the observed frequencies are extremely precise.
  \item The low-$m$ mixed modes and high-$m$ fundamental modes have yet to be addressed jointly in a single Saturn model.
  \item The seismology, gravity, and magnetic data from \cassini have not been addressed jointly in a single Saturn model. For a start, the normal mode eigenfrequencies and zonal gravity harmonics should be {fit} simultaneously to provide better-constrained Saturn interior models. This will significantly diminish degeneracies inherent to each dataset taken in isolation.
  \item {The generation of Saturn's magnetic field is not understood in the context of a thick stable stratification occupying the deepest parts of the electrically conductive interior. }
\end{enumerate}

Finally, the most basic puzzle that remains is how normal mode oscillations in giant planets are excited in the first place. Turbulent convection, the mechanism powering the solar oscillations, is almost certainly ineffective in the{ vastly} dimmer Jupiter and Saturn. Some imaginative ideas have appealed to rock storms \cite{2018Icar..306..200M} and ancient giant impacts \cite{2019ApJ...881..142W}, but neither theory provides a completely satisfactory fit to the Saturn ring wave amplitudes reported by \citeA{2019AJ....157...18H}. In this arena as in the others, it appears that theory has some catching up to do.

\acknowledgments
I wish to thank Francis Nimmo, David Al-Attar, and an anonymous reviewer for their helpful comments on this manuscript.
I also gratefully acknowledge support from the Division of Geological and Planetary Sciences at Caltech. Figures \ref{fig.harmonics} and \ref{fig.mode_wave_schematic} were created with the aid of code by Keaton J. Bell (\url{https://github.com/keatonb/sphericalharmonics}). No data were generated in the course of this work.

\bibliography{kronoseis}

\end{document}